\documentclass[conference]{IEEEtran}
\IEEEoverridecommandlockouts

\usepackage{amsmath}
\usepackage[utf8]{inputenc} 
\usepackage{amsmath, amssymb, amsfonts} 
\usepackage{textcomp} 

\usepackage{graphicx} 
\usepackage[margin=1in]{geometry} 
\usepackage{multicol} 
\usepackage{makecell, multirow, tabularray, array} 

\usepackage{listings} 
\usepackage{xcolor} 

\usepackage{tikz} 
\usetikzlibrary{matrix, shapes, arrows, positioning, chains, calc} 
\usepackage{pgfplots} 
\pgfplotsset{compat=1.17} 

\usepackage{lipsum} 
\usepackage{cite} 
\usepackage{algorithmic} 

\definecolor{dkgreen}{rgb}{0,0.6,0}
\definecolor{gray}{rgb}{0.5,0.5,0.5}
\definecolor{mauve}{rgb}{0.58,0,0.82}
\definecolor{color1}{RGB}{175, 225, 218}
\definecolor{color2}{RGB}{232, 191, 125}
\definecolor{color3}{RGB}{168, 211, 238}
\definecolor{color4}{RGB}{97, 76, 151}

\tikzstyle{startstop} = [rectangle, text width=3cm, minimum height=1cm, text centered, draw=black, fill=red!30]
\tikzstyle{io} = [trapezium, trapezium left angle=70, trapezium right angle=110, text width=3cm, minimum height=1cm, text centered, draw=black, fill=blue!30]
\tikzstyle{process} = [rectangle, text width=3cm, minimum height=1cm, text centered, draw=black, fill=orange!30]
\tikzstyle{decision} = [diamond, text width=3cm, minimum height=1cm, text centered, draw=black, fill=green!30]
\tikzstyle{arrow} = [thick, ->, >=stealth]

\def\BibTeX{{\rm B\kern-.05em{\sc i\kern-.025em b}\kern-.08em T\kern-.1667em\lower.7ex\hbox{E}\kern-.125emX}}

\begin{document}

\newcommand{\ket}[1]{\left| #1 \right>}

\title{Implementation Guidelines and Innovations in Quantum LSTM Networks\thanks{}}

\author{

\IEEEauthorblockN{Yifan Zhou*}
\IEEEauthorblockA{
\textit{University of California, Los Angeles}\\
Los Angeles, United States of America\\
yzhou05@ucla.edu}
~\\
\and
\IEEEauthorblockN{Chong Cheng Xu}
\IEEEauthorblockA{
\textit{BASIS International School Guangzhou}\\
Guangzhou, China \\
chongcheng.xu12245-bigz@basischina.com}
~\\
\and
\IEEEauthorblockN{Mingi Song}
\IEEEauthorblockA{
\textit{BASIS International School Guangzhou}\\
Guangzhou, China \\
mingi.song16149-bigz@basischina.com}
~\\
\and
\IEEEauthorblockN{Yew Kee Wong}
\IEEEauthorblockA{
\textit{Hong Kong Chu Hai College}\\
Hong Kong, China \\
ericwong@chuhai.edu.hk}
\and
\IEEEauthorblockN{Kangsong Du}
\IEEEauthorblockA{
\textit{University of California, Davis}\\
Davis, California \\
kevdu@ucdavis.edu}
}

\maketitle


\begin{abstract}
The rapid evolution of artificial intelligence has driven interest in Long Short-Term Memory (LSTM) networks for their effectiveness in processing sequential data. However, traditional LSTMs are limited by issues such as the vanishing gradient problem and high computational demands. Quantum computing offers a potential solution to these challenges, promising advancements in computational efficiency through the unique properties of qubits, such as superposition and entanglement. This paper presents a theoretical analysis and an implementation plan for a Quantum LSTM (qLSTM) model, which seeks to integrate quantum computing principles with traditional LSTM networks. While the proposed model aims to address the limitations of classical LSTMs, this study focuses primarily on the theoretical aspects and the implementation framework. The actual architecture and its practical effectiveness in enhancing sequential data processing remain to be developed and demonstrated in future work.
\end{abstract}



\section{Introduction}
\subsection{Review of LSTM: structure, functionality, and limitations}

Recurrent neural networks (RNNs) process sequential data through looping connections, but they often encounter the vanishing gradient problem. This issue arises when gradients, which are used for adjusting weights, exponentially decrease in magnitude as they propagate backward through the network's layers.\cite{b1} Fortunately, the LSTM (Long Short-Term Memory) is a specific type of recurrent neural network that applies a complex memory recall model to assuage the vanishing gradient problems in traditional recurrent neural networks. \cite{b2} Specifically, its unique architecture of utilizing the forget, input, and output gates to update the cell states and hidden states initiates the rise of diverse functionalities such as natural language processing, time series prediction, and speech recognition. \cite{b2}  Although LSTM has a rich array of usage in today’s world, its’ limits remain with the greater computational complexity of multiple gates, the continued presence of vanishing gradient problems, and challenges associated with tuning hyper-parameters. 

\subsection{Introduction to quantum computing: basic principles relevant to machine learning}

In the contemporary digital era, most machine learning models, including LSTMs, operate on classical computing principles, utilizing binary bits (0s and 1s) and logical gates. However, the rise of new technology leads to quantum computing, which applies qubits that can be simultaneously in both 0 and 1 states, representing multiple states at once. This principle, called superposition, enables quantum algorithms to process multiple calculations in parallel, increasing their efficiency. \cite{b3} Furthermore, the principle of quantum entanglement could address traditional LSTMs' limitations in context understanding, as it allows for direct connections between distant data points or qubits.\cite{b3} Thus, the use of quantum gates implementing the two principles in quantum circuits leads to various quantum machine learning models including quantum neural networks, quantum support vector machines, quantum principal component analysis, and quantum Boltzmann machines.\cite{b4,b5,b6}

\subsection{Rationale for integrating quantum computing with LSTMs}

There are a variety of reasons to incorporate quantum computing into LSTM models. From one aspect, replacing classical bits with qubits incorporating superposition expands the memory capacity of LSTM models as the length sequences increase.\cite{b3} From another aspect, quantum machine learning algorithms such as quantum support vector machines and quantum k-means clustering can be more effective data analysis that facilitates the LSTM model to more accurately decide whether to forget or pass on the data set. \cite{b4,b7} Furthermore, the lack of context understanding in traditional LSTM models can be tackled through the principles of entanglement in quantum computing, as one data set or qubit can be directly linked to another data set or qubit despite their overall distances. This means that through quantum LSTMs, the current data set can successfully trace back to its linking data set in the past. Lastly, quantum principal component anal
ysis (PCA) or quantum-based feature selection methods can indirectly solve the challenges associated with hyperparameter tuning by facilitating feature selection by reducing data’s dimensionality.\cite{b8,b9}

Outline of the paper's structure and main contributions

\section{Quantum Computing Preliminaries}
Basic knowledge of quantum computing preliminaries such as Hardmard gates and qubits is required to understand how LSTM can be applied to a quantum setting. 

\subsection{Quantum bits}

Currently, classical computing systems operate on classical bits, two distinct states, shown by binary numbers 0 and 1. A quantum bit, on the other hand, is a two-level quantum system defined by a two-dimensional complex Hilbert space. This means that before the qubit is revealed, it can be in an interchangeable state of both 1s and 0s based on probability.\cite{b3} In addition, a pair of quantum states can correspond to the classical bits of 0 and 1. 
\begin{equation}
    |0\rangle \equiv \left [\begin{array}{cc} 1 \\ 0\end{array} \right]\mbox{ , }|1\rangle \equiv \left[\begin{array}{cc} 0 \\ 1\end{array} \right ]
\end{equation}

The two quantum states construct the basic computational vectors. Resulting from the superposition principle to add vectors, a state of the quantum bit can be written in the form of:
\begin{equation}
|\phi\rangle = \alpha|0\rangle + \beta|1\rangle
\end{equation}

in which the variables alpha and beta are complex numbers and obey the normalization condition:
\begin{equation}
\left| \alpha \right|^2 + \left| \beta \right|^2 = 1
\end{equation}

In conclusion, a quantum bit differs significantly in its overall structure since it is a vector space described by variables, while classical bits are just 1s and 0s.

\subsection{Quantum gates}
While classical logic gates focus on deterministic operations like AND, OR, and NOR, implementing sequential computation, quantum logic gates implement parallel computations using reversible unitary transformations. Applying quantum logic gates as they manipulate the characteristics of superposition and entanglement is crucial for the operation of advanced quantum circuits. 

\paragraph{Single qubit gates}
A single qubit gate that operates on a single qubit described by vectors $|0\rangle$ and $|1\rangle$ is defined by a 2 x 2 unitary matrix. \cite{b3}

\begin{enumerate}
    \item \textbf{Hadamard Gate}: The Hadamard Gate transforms the basic vectors 0 and 1 into vectors + and -, applying superposition principles. This means that the Hadamard gate maps the $|0\rangle$ state to an equal superposition of $|0\rangle$ and $|1\rangle$, and the $|1\rangle$ state to an equal superposition of $|0\rangle$ and -$|1\rangle$. \cite{b3} Furthermore, the Hadamard gate is self-inverse, meaning that applying it twice brings the qubit back to its original state. \cite{b3}

    \item \textbf{Phase Gate}: The phase gate presents a phase shift of $\pi/2$ radians to the $|1\rangle$ state while leaving the $|0\rangle$ state unchanged.\cite{b3} Therefore, $|1\rangle$ turns to -$|1\rangle$.

    \item \textbf{T Gate}:The T gate introduces a phase shift of $\pi/4$ radians to the $|1\rangle$ state while leaving the $|0\rangle$ state unchanged. \cite{b3} Accordingly, $|1\rangle$ gets transformed into $e^{i\frac{\pi}{4}}|1\rangle$.

    \item\textbf{Rotation Gate}: $R\scriptsize x$, $R\scriptsize y$, and $R\scriptsize z$ are the rotation gates for a single qubit. Rotations are carried out based on the $x$, $y$, and $z$ axes in the Hilbert space, overall transforming the quantum bit. The matrix representations of the three gates consist of a variety of trigonometric, imaginary, and exponential functions. 
\begin{itemize}
\item \textbf{$R\scriptsize x$}: The $R\scriptsize x$ gate applies a rotation of $\Theta$ radians around the x-axis, changing the probability amplitudes of the $|0\rangle$ and $|1\rangle$ states.\cite{b3}

\item \textbf{$R\scriptsize y$}: The $R\scriptsize y$ gate causes a rotation of $\Theta$ radians around the y-axis, altering the quantum state's probability amplitudes.\cite{b3}

\item \textbf{$R\scriptsize z$}: The $R\scriptsize z$ gate introduces a phase shift of $\Theta$ radians to the $|1\rangle$ state while leaving the $|0\rangle$ state unchanged.\cite{b3}
\end{itemize}
    
\end{enumerate}

\paragraph{Controlled-NOT gate, Controlled SWAP gate, Toffoli Gate}
These gates perform functions with inputs from multiple qubits. The primary purpose of these gates is to establish the creation of complex entangled states and the manipulation of these states to perform quantum algorithms. 
\begin{enumerate}

\item\textbf{Controlled Not Gate}: The Controlled Not gate (CNOT) is a typical two qubit gate that applies the principles of entanglement.\cite{b3} This operates on a control qubit and a target qubit. If the control qubit is 1, the target qubit gets rotated by $\pi$ radians, flipping the value of the target qubit.\cite{b3} If the control qubit is 0 in the beginning, the target qubit remains constant. 

\item \textbf{Controlled-SWAP Gate}: The Controlled-SWAP gate(CSWAP) operates on two target qubits and one control qubit. If the control qubit is set to one, the value of the two target qubits gets swapped. \cite{b3} Else if the control qubit has 0 as its initial state, the target qubits remain unchanged.

\item \textbf{Toffoli Gate}: The Toffoli gate is a three-qubit gate that performs the NOT operation on the target qubit when both control qubits are in the state $|1\rangle$. \cite{b3} Otherwise, it leaves the target qubit unchanged. The Toffoli gate can be considered a generalized version of the CNOT gate.
\end{enumerate}

\subsection{Quantum circuits and measurement}
Quantum circuits utilize a variety of quantum logic gates to achieve various quantum computations. Measurement acts as the final step in a circuit, in which the quantum states are observed, resulting the state to "collapse" or to become a specific state of 0 or 1. In order to measure the final values of quantum bits, common operators are $|0\rangle \langle 0|$ and $|1\rangle \langle 1|$. The probability of measuring a certain state is given by corresponding to operator P. \cite{b3}

\paragraph{Ansatz quantum circuits}
Ansatz quantum circuit is a parameterized trial circuit that begins with an initial guess and implements variational quantum algorithms to operate numerous times in order to approach a solution to a problem.\cite{b50} This is typically used to approximate a specific mathematical operation or function.  To achieve this, an Ansatz is usually composed of rotational quantum gates in the x, y, and z directions.\cite{b50} After that, the parameters of the Ansatz quantum circuit are altered to minimize or maximize the cost function, decreasing the gap between the actual solution and the calculated solution.\cite{b50} Below are the steps for how Ansatz quantum circuits work:

\begin{enumerate}
\item\textbf{Definition}: The Ansatz quantum circuit has to be defined and initialized with adjustable parameters like $\theta$ and quantum logic gates such as the rotational gates. \cite{b50}
\item\textbf{Initialization}: Classical data has to be converted to a quantum state by entering through quantum logic gates like Hadamard gates. The Hadamard gate converts $|1\rangle$ and $|0\rangle$ into vectors with equal probability of both $|1\rangle$ and $|0\rangle$. Then, the quantum data is inputted into the circuit to initialize the circuit. 
\item\textbf{Measurement and Cost Function}: After applying the quantum circuit, the quantum data is measured. Furthermore, the cost function is set up to measure the differences between the target values and the measured values. \cite{b50}
\item\textbf{Parameter Optimization}: Parameters are adjusted to minimize the cost functions.\cite{b50}
\end{enumerate}

\section{Quantum LSTM: Architecture and Design}

The Quantum LSTM (qLSTM) model represents a cutting-edge fusion of quantum computing principles and Long Short-Term Memory (LSTM) networks. The main goal of the qLSTM is to harness the distinctive properties of quantum mechanics, such as superposition, entanglement, and quantum interference, to overcome the inherent limitations of classical LSTMs. This section provides an in-depth explanation of the qLSTM model's architecture, detailing each component's design, purpose, and functionality, and illustrating how they work together to enhance the model's performance.

\subsection{Overview of qLSTM}

The qLSTM model consists of several interconnected components that collectively process sequential data in a highly efficient and effective manner. The process begins with the Quantum Embedding Layer, which encodes classical input data into quantum states. This encoded data is then passed to the Quantum Memory Cell, where quantum gates perform state updates and maintain information. The Classical-Quantum Interface plays a critical role in facilitating communication between the classical and quantum components, encoding classical data into quantum states and decoding quantum states back into classical data. Each component has a specific role and is designed to work in harmony with the others, leveraging the strengths of quantum computing to improve performance and efficiency.

\subsection{Step-by-Step Process}

\begin{enumerate}
    \item \textbf{Classical Data Input}: The process begins with the input of classical data into the model. This data can be in various forms, such as time series, text, or any other sequential data.
    
    \item \textbf{Quantum Embedding Layer}: The classical data is transformed into quantum states using the Quantum Embedding Layer. This layer uses quantum gates, particularly the Hadamard gate, to encode classical bits into quantum superposition states. This step is crucial as it sets up the data for quantum processing, enabling parallelism and richer data representation.
    
    \item \textbf{State Initialization}: In the Quantum Memory Cell, the initial states of the qubits are prepared. These states represent the initial cell and hidden states, analogous to classical LSTM networks.
    
    \item \textbf{Quantum Gates Operation}: The Quantum Memory Cell updates the cell and hidden states using a series of quantum gates. These gates include the Hadamard, CNOT, Phase, and Quantum Fourier Transform (QFT) gates, which manipulate the quantum states to perform memory updates, maintain information, and facilitate state transitions.
    
    \item \textbf{State Update}: The updated quantum states (cell state $\ket{c_t}$ and hidden state $\ket{h_t}$) are computed using the quantum analogs of the forget, input, and output gates. These updates leverage the principles of superposition and entanglement to enhance memory retention and context understanding.
    
    \item \textbf{Classical-Quantum Interface}: The processed quantum states are then decoded back into classical data through the Classical-Quantum Interface. This interface ensures that the quantum computations are effectively utilized in the classical domain by converting the quantum states into a format that can be further analyzed and used for decision-making.
    
    \item \textbf{Output Generation}: The final step involves generating the output based on the processed data. This output can be predictions, classifications, or any other relevant results, depending on the application.
\end{enumerate}

Each of these steps is designed to enhance the model's ability to handle complex data sequences, leveraging the unique properties of quantum computing to overcome the limitations of classical LSTMs. The following subsections provide detailed explanations of each component within the qLSTM architecture, illustrating their specific roles and contributions to the overall system.

\subsection{Quantum Embedding Layer}
The Quantum Embedding Layer is the initial stage of the qLSTM model, responsible for encoding classical input data into quantum states. This transformation is crucial, as it enables the model to leverage quantum parallelism and the enhanced state space of qubits. The Quantum Embedding Layer effectively transforms the input space, providing a richer and more complex representation of the data compared to classical embeddings.

\paragraph{Transformation Process}
The transformation of a classical bit $\ket{x}$ using the Hadamard gate is represented as:
\begin{equation}
\ket{x} \rightarrow H\ket{x} = \frac{1}{\sqrt{2}}(\ket{0} + \ket{1})
\end{equation}
In this superposition state, the qubit exists simultaneously in both 0 and 1 states, allowing parallel processing of multiple states, which significantly enhances the data processing capacity of the model.

\paragraph{Design Choice}
The Hadamard gate is chosen for its ability to create a superposition state, which is fundamental in quantum computing for exploring multiple possibilities simultaneously. By encoding classical data into quantum states, the embedding layer lays the groundwork for exploiting the vast computational potential of quantum mechanics.

\paragraph{Integration with the System}
The Quantum Embedding Layer works in conjunction with the subsequent Quantum Memory Cell by providing a quantum representation of the input data, which is then processed by the quantum gates to update the memory cell states.

\subsection{Quantum Memory Cell Implementation}
The Quantum Memory Cell is the core innovation of the qLSTM model, analogous to the cell state in classical LSTMs. Unlike classical LSTM cells, which are constrained by issues like the vanishing gradient and computational inefficiency, the Quantum Memory Cell leverages the principles of quantum computing to overcome these limitations. This subsection delves into the specific implementation details of the qLSTM elements, offering a clear understanding of how quantum operations are integrated into the model.

\paragraph{State Update Mechanism}
The state update mechanism in the Quantum Memory Cell operates through quantum operations that directly correspond to the functionalities of classical LSTM gates. The update rules for the cell state $\ket{c_t}$ and hidden state $\ket{h_t}$ are expressed as:
\begin{equation}
\ket{c_t} = f_t \circ \ket{c_{t-1}} + i_t \circ \tilde{c}_t
\end{equation}
\begin{equation}
\ket{h_t} = o_t \circ \tanh(\ket{c_t})
\end{equation}
where $f_t$, $i_t$, and $o_t$ are quantum implementations of the forget, input, and output gates, respectively. Each of these gates is realized through a combination of quantum operations, specifically tailored to maintain and manipulate quantum states within the Quantum Memory Cell.

\paragraph{Quantum Gate Implementation in qLSTM}
Instead of a generic description of quantum gates, this section details how each gate is implemented within the qLSTM architecture:

\begin{itemize}
    \item \textbf{Quantum Forget Gate ($f_t$)}: This gate is responsible for modulating the retention of previous cell state information. It is implemented by applying a series of controlled rotations (e.g., controlled-$R_z$) based on the quantum input, which determines the degree to which the previous cell state $\ket{c_{t-1}}$ is preserved.
    
    \item \textbf{Quantum Input Gate ($i_t$)}: The input gate controls the incorporation of new information into the cell state. It utilizes a combination of controlled Hadamard operations and phase shifts to prepare the input qubits, followed by a controlled-NOT (CNOT) operation to entangle these with the current state, enabling the effective integration of new data into the memory cell.
    
    \item \textbf{Quantum Output Gate ($o_t$)}: The output gate governs the information passed to the hidden state $\ket{h_t}$. This gate is implemented through a quantum circuit that applies a Hadamard gate to create superpositions, followed by a series of controlled phase gates to extract the relevant information based on the current cell state $\ket{c_t}$. The result is then measured to produce the final output.
\end{itemize}

\paragraph{Data Decoding and Probability Sampling}
A critical component of the qLSTM's functionality is the process of decoding the quantum data into a classical form that can be interpreted for downstream tasks. This involves sampling from the probability distribution of the qubit states, a step that is crucial for understanding the outcomes of the quantum computations.

The decoding process is performed as follows:
\begin{enumerate}
    \item \textbf{Measurement of Qubits:} The qubits corresponding to the cell state $\ket{c_t}$ and the hidden state $\ket{h_t}$ are measured in the computational basis. This measurement collapses the qubit superpositions, yielding a classical bitstring that represents the state of the system at time $t$.
    
    \item \textbf{Sampling from the Probability Distribution:} After measurement, the resulting bitstrings represent samples from the underlying probability distribution of the quantum states. To generate meaningful predictions, multiple measurements are performed, allowing for the estimation of this distribution.
    
    \item \textbf{Post-Processing:} The sampled bitstrings are then processed through classical post-processing steps, such as aggregating the results or applying classical machine learning algorithms, to interpret the output in the context of the sequence learning task.
\end{enumerate}

\paragraph{Design Rationale and System Integration}
The integration of quantum gates in the qLSTM architecture is not arbitrary but driven by the need to effectively simulate the functionalities of classical LSTM components in a quantum framework. By carefully selecting and applying quantum operations, the qLSTM model achieves a balance between retaining the power of classical LSTM networks and exploiting the advantages of quantum computing.

The Quantum Memory Cell, as part of this integrated system, plays a pivotal role in the overall architecture by receiving quantum-encoded inputs, performing state updates, and passing the updated states for further processing through the Classical-Quantum Interface. This careful orchestration of quantum operations ensures that the qLSTM model can handle sequential data tasks with enhanced efficiency and effectiveness.

\subsection{Classical-Quantum Interface}
The classical quantum Interface is a crucial component that ensures seamless communication between the classical and quantum parts of the qLSTM model. This interface handles the encoding of classical input data into quantum states and the decoding of quantum states back into classical data for interpretation and further analysis.

\paragraph{Data Encoding}
Classical data $x$ is encoded into a quantum state $\ket{\psi}$:
\begin{equation}
\ket{\psi} = \sum_{i=0}^{N-1} \alpha_i \ket{i}
\end{equation}
where $\alpha_i$ are amplitude coefficients derived from the classical data.

\paragraph{Data Decoding}
Quantum measurements are performed to extract classical information from the quantum state:
\begin{equation}
P(\ket{i}) = |\alpha_i|^2
\end{equation}
This probability distribution is used to reconstruct the classical data.

\paragraph{Design Considerations}
The interface employs quantum measurement techniques to extract meaningful information from the quantum states, ensuring that quantum computations are effectively utilized in the classical domain. This component is essential for maintaining the integrity and usability of the processed data.

\paragraph{Integration with the System}
The Classical-Quantum Interface acts as the bridge between the Quantum Memory Cell and the classical output layers, ensuring that the processed quantum information is converted back into a format that can be used for further analysis and decision-making.

\subsection{Quantum Circuits}
Ansatz quantum circuits are essential to the qLSTM model, approximating mathematical functions such as $\sigma(x)$ and $\tanh(x)$ based on a quantum context.\cite{b50} These circuits operate multiple times with the goal of minimizing errors between the target value and the measured value. \cite{b50} For example, the mean squared error function in a parametric quantum circuit can be shown by:
\begin{equation}
\text{MSE}(\theta) = \frac{1}{N} \sum_{i=1}^{N} \left( \langle \phi_i(\theta) | Z | \phi_i(\theta) \rangle - y_i \right)^2
\end{equation}

\paragraph{Design Considerations} The circuit takes input from previous output qubits and current input qubits. Thus, a circuit with multiple adjustable parameters is required to operate based on certain mathematical functions to determine the probability of forgetting or preserving certain information. 

\paragraph{Integration with the System} Ansatz quantum circuits are employed to replace the classical forget, input, and output gates with quantum algorithms that can successfully approximate the mathematical functions after multiple trials.

\section{Innovative Features}

\subsection{Enhanced Memory Capacity through Superposition}
The qLSTM model leverages quantum superposition to significantly enhance memory capacity. Traditional LSTM models face challenges with long sequences due to the vanishing gradient problem. By using qubits in superposition, the qLSTM can maintain and process exponentially larger states, effectively mitigating the vanishing gradient issue and improving long-term dependency handling.

\paragraph{Mathematical Representation}
A quantum state representing superposition can be expressed as:
\begin{equation}
\ket{\psi} = \sum_{i=0}^{2^n-1} \alpha_i \ket{i}
\end{equation}
This allows the model to retain relevant information over extended sequences without the degradation typically seen in classical models.

\paragraph{Integration with the System}
The enhanced memory capacity provided by superposition allows the Quantum Memory Cell to store and process information more efficiently, improving the overall performance of the qLSTM model in handling long-term dependencies.

\subsection{Context Understanding through Entanglement}
Quantum entanglement allows the qLSTM to capture and utilize long-range dependencies more effectively than classical LSTMs. Entangled qubits can maintain context across distant elements in a sequence, significantly enhancing the model's performance in tasks requiring deep contextual understanding.

\paragraph{Entangled States}
Entanglement can be represented by an entangled state, such as the Bell state:
\begin{equation}
\ket{\Phi^+} = \frac{1}{\sqrt{2}} (\ket{00} + \ket{11})
\end{equation}
This property allows for immediate and robust context propagation, improving performance in natural language processing and complex sequence prediction.

\paragraph{Integration with the System}
By leveraging entanglement, the Quantum Memory Cell can maintain strong correlations between distant data points, enhancing the model's ability to understand and process context over long sequences.

\subsection{Parallelism and Computational Efficiency}
Quantum parallelism enables the qLSTM to perform multiple operations simultaneously, significantly boosting computational efficiency. This reduces the computational overhead and accelerates training and inference phases, making the qLSTM model more scalable and efficient for large-scale applications.

\paragraph{Quantum Parallelism}
Quantum gate operations on superposed states can be represented as:
\begin{equation}
U\ket{\psi} = \sum_{i=0}^{2^n-1} U\ket{\alpha_i}
\end{equation}
This parallel processing capability allows the qLSTM to handle larger datasets and more complex models efficiently.

\paragraph{Integration with the System}
The parallelism inherent in quantum computing enables the qLSTM model to perform more calculations in less time, enhancing its scalability and making it suitable for real-time processing applications.

\subsection{Quantum-Based Feature Selection and Dimensionality Reduction}
The qLSTM incorporates quantum principal component analysis (PCA) and other quantum-based feature selection methods to address hyperparameter tuning and data dimensionality challenges. By leveraging quantum algorithms, the qLSTM can reduce input data dimensionality more effectively, enhancing performance and simplifying the tuning process.

\paragraph{Quantum PCA}
The quantum PCA algorithm involves:
\begin{enumerate}
    \item \textbf{State Preparation}: Prepare the quantum state representing the data matrix $X$:
    \begin{equation}
    \ket{\psi} = \frac{1}{\|X\|} \sum_{i,j} X_{ij} \ket{i} \ket{j}
    \end{equation}
    \item \textbf{Phase Estimation}: Apply phase estimation to obtain eigenvalues and eigenvectors of the covariance matrix $X^TX$.
    \item \textbf{Measurement}: Measure the eigenvalues and select corresponding eigenvectors as principal components.
\end{enumerate}

\paragraph{Integration with the System}
Quantum-based feature selection improves the qLSTM model's ability to generalize from training data to unseen examples by reducing data dimensionality, thereby enhancing predictive accuracy and computational efficiency.

\section{Implementation Guidlines for Quantum LSTM}

The implementation of the Quantum LSTM (qLSTM) model involves several stages, each crucial for ensuring the effective integration of quantum computing principles with the LSTM framework. This section outlines the detailed steps required to implement the qLSTM model, covering hardware setup, algorithm development, circuit design, and performance evaluation. The implementation plan is designed to ensure that each component of the qLSTM model is meticulously developed, tested, and validated, paving the way for practical applications and further advancements in the field of quantum-enhanced sequential modeling.

\subsection{Hardware Setup}

Implementing the qLSTM model necessitates access to quantum hardware capable of executing complex quantum operations with high fidelity. The hardware setup for this implementation involves several critical considerations, beyond just the selection of a quantum platform. This subsection provides a detailed rationale for choosing specific quantum hardware and discusses the importance of implementation nuances that directly impact the effectiveness of the proposed algorithm.

\paragraph{Selection of Quantum Hardware Platform}
The choice of quantum hardware platform is crucial and extends beyond the mere availability of qubits or gate fidelity. While superconducting quantum processors, such as those provided by IBM Qiskit, Google Quantum Processor, or Rigetti Quantum Cloud Services, are popular due to their advanced development and accessibility, other platforms should also be considered depending on the specific requirements of the qLSTM model. For instance, trapped ion quantum processors offer longer coherence times and potentially lower gate error rates, while photonic quantum processors can facilitate faster gate operations and better scalability for specific tasks.

The selection criteria for the quantum hardware should include:
\begin{itemize}
    \item \textbf{Qubit Quality and Gate Fidelity}: The coherence time, gate error rates, and cross-talk between qubits are critical factors that influence the reliability of quantum computations.
    \item \textbf{Scalability}: The ability of the quantum processor to scale up in terms of qubit count without significant degradation in performance is essential for implementing more complex qLSTM architectures.
    \item \textbf{Operational Speed}: The execution speed of quantum operations, especially in comparison to the classical components, affects the overall performance of the hybrid quantum-classical system.
    \item \textbf{Integration Capability}: The ease with which the quantum platform can be integrated with classical computing systems for seamless data exchange and processing.
\end{itemize}
While superconducting QPUs are a strong candidate due to their current dominance in the field, alternatives like trapped ion or photonic quantum processors might be considered depending on the specific use case and available infrastructure \cite{b1, b2, b3}.

\paragraph{Quantum Processor Configuration}
Once a suitable quantum platform is selected, configuring the quantum processor to support the qLSTM model is a critical step. This configuration involves:

\begin{itemize}
    \item \textbf{Initialization of the Quantum Environment}: This step includes setting up the quantum computing environment, which may involve calibrating qubits, optimizing gate sequences, and setting the initial quantum states necessary for the qLSTM operations.
    \item \textbf{Allocation of Qubits}: Qubit allocation must be done strategically to minimize error propagation and maximize the efficiency of quantum gate operations. This includes considerations for qubit connectivity (e.g., nearest-neighbor couplings) and the layout of qubits to ensure effective implementation of quantum operations.
    \item \textbf{Gate Setup and Calibration}: The specific quantum gates required by the qLSTM model, such as controlled-NOT (CNOT) gates, phase gates, and quantum Fourier transforms (QFT), must be carefully calibrated and optimized for the chosen hardware. The gate fidelity and the noise characteristics of the hardware will directly influence the model's accuracy and robustness \cite{b4, b5}.
\end{itemize}

\paragraph{Integration with Classical Systems}
Effective implementation of the qLSTM model requires seamless integration between quantum and classical systems. This integration involves:

\begin{itemize}
    \item \textbf{Communication Protocols}: Establishing efficient communication protocols between the quantum and classical components is essential for real-time data exchange. This might involve low-latency networking solutions or specialized software interfaces that can handle the hybrid computational workflow.
    \item \textbf{Data Transfer and Processing}: The classical system must be capable of rapidly processing the output from quantum operations, such as the results of qubit measurements, and feeding this data back into the quantum system as needed. This feedback loop is crucial for implementing iterative processes within the qLSTM model.
    \item \textbf{Hybrid Algorithm Optimization}: The overall effectiveness of the qLSTM algorithm is heavily dependent on how the quantum and classical systems are synchronized. This synchronization ensures that the computational workload is optimally distributed and that the inherent strengths of both quantum and classical processing are fully leveraged \cite{b6, b7}.
\end{itemize}

\paragraph{Assessment of Effectiveness}
The effectiveness of the proposed qLSTM model cannot be fully evaluated without a thorough understanding of the hardware-specific implementation details. Simply conceptualizing the hardware setup is insufficient; the success of the qLSTM algorithm hinges on how quantum operations are executed on the chosen platform. For this reason, a detailed analysis, including circuit diagrams and specific implementation strategies, is essential to assess the practical viability and performance of the model.

By providing a more granular description of the hardware setup, including the rationale behind platform selection and the integration with classical systems, this subsection aims to offer the necessary insights for accurately evaluating the qLSTM model's potential in real-world applications.

\subsection{Algorithm Development}

Developing quantum algorithms is a critical step in implementing the qLSTM model. These algorithms are responsible for state initialization, gate operations, and state updates. The steps involved are:

\begin{enumerate}
    \item \textbf{State Initialization Algorithms}: Develop algorithms to initialize the quantum states representing the initial cell and hidden states. These algorithms should ensure that the qubits are prepared accurately for subsequent operations \cite{b8}.
    \item \textbf{Quantum Gate Operations}: Implement algorithms for the Hadamard, CNOT, Phase, and Quantum Fourier Transform (QFT) gates. These algorithms should efficiently perform the required quantum operations on the qubits \cite{b9}.
    \item \textbf{State Update Mechanisms}: Develop mechanisms to update the cell and hidden states based on the quantum analogs of the forget, input, and output gates. These updates should leverage superposition and entanglement to maintain information and context \cite{b10}.
\end{enumerate}

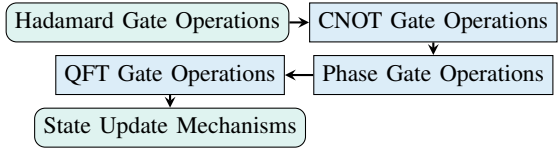
\begin{figure}
    \centering
   
        \begin{tikzpicture}[node distance=1cm, font = \small]
        \centering
        \tikzstyle{startstop} = [rectangle, rounded corners, minimum width=1.5cm, minimum height=0.5cm, text centered, draw=black, fill=color1!40, font=\small]
        \tikzstyle{io} = [trapezium, trapezium left angle=70, trapezium right angle=110, minimum width=1.5cm, minimum height=0.5cm, text centered, draw=black, fill=color2!40, font=\small]
        \tikzstyle{process} = [rectangle, minimum width=1.5cm, minimum height=0.5cm, text centered, draw=black, fill=color3!40, font=\small]
        \tikzstyle{decision} = [diamond, minimum width=1.5cm, minimum height=0.5cm, text centered, draw=black, fill=color4!40, font=\small]
        \tikzstyle{arrow} = [thick,->,>=stealth]
     
\node (hadamard) [startstop] {Hadamard Gate Operations};
\node (cnot) [process, right of=hadamard, xshift = 2.8cm] {CNOT Gate Operations};
\node (phase) [process, below of=cnot, yshift = 0.3cm] {Phase Gate Operations};
\node (qft) [process, left of=phase, xshift = -2.5cm] {QFT Gate Operations};
\node (update) [startstop, below of=qft, yshift = 0.3cm] {State Update Mechanisms};

\draw [arrow] (hadamard) -- (cnot);
\draw [arrow] (cnot) -- (phase);
\draw [arrow] (phase) -- (qft);
\draw [arrow] (qft) -- (update);

\end{tikzpicture}

    \caption{Figure Showing Algorithm Development}
    \label{Algorithm Development}
\end{figure}

\subsection{Quantum Circuit Design}

Designing the quantum circuits is essential for the practical realization of the qLSTM model. The circuit design process includes:
    \begin{figure} [ht]
        \centering

        \begin{tikzpicture}[node distance=1cm, font = \small]
        \centering
        \tikzstyle{startstop} = [rectangle, rounded corners, minimum width=1.5cm, minimum height=0.5cm, text centered, draw=black, fill=color1!40, font=\small]
        \tikzstyle{io} = [trapezium, trapezium left angle=70, trapezium right angle=110, minimum width=1.5cm, minimum height=0.5cm, text centered, draw=black, fill=color2!40, font=\small]
        \tikzstyle{process} = [rectangle, minimum width=1.5cm, minimum height=0.5cm, text centered, draw=black, fill=color3!40, font=\small]
        \tikzstyle{decision} = [diamond, minimum width=1.5cm, minimum height=0.5cm, text centered, draw=black, fill=color4!40, font=\small]
        \tikzstyle{arrow} = [thick,->,>=stealth]

\node (embed) [startstop] {Quantum Embedding Circuit};
\node (memory) [process, below of=embed] {Quantum Memory Circuit};
\node (ansatz) [process, below of=memory] {Ansatz Circuit};
\node (measure) [startstop, below of=ansatz] {Measurement Circuit};

\draw [arrow] (embed) -- (memory);
\draw [arrow] (memory) -- (ansatz);
\draw [arrow] (ansatz) -- (measure);

\end{tikzpicture}
        \caption{Figure Showing Circuit Design}
        \label{Circuit Design}
\end{figure}
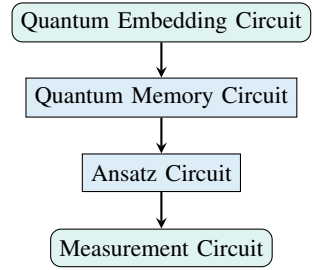
\begin{enumerate}
    \item \textbf{Quantum Embedding Circuit}: Design the circuit that transforms classical inputs into quantum states using the Hadamard gate. This circuit should ensure accurate encoding of data into superposition states \cite{b11}.
    \item \textbf{Quantum Memory Circuit}: Develop the circuit that implements the quantum memory cell, incorporating Hadamard, CNOT, Phase, and QFT gates to update and maintain quantum states. This circuit must efficiently handle state transitions and information retention \cite{b12}.
    \item \textbf{Ansatz Circuit}: Develop and train the circuit to approximate mathematical functions applied to quantum input. Advance the classical logic forget, input, and output gates with quantum logic forget, input, and output gates through implementing rotational $Rx, Ry, Rz$ gates with adjustable parameters. \cite{b50}
    \item \textbf{Measurement Circuit}: Create the circuit for measuring quantum states and converting them back into classical data. This circuit is crucial for decoding quantum information and integrating it with classical processing \cite{b13}.

\end{enumerate}

\begin{figure*}[!ht]
    \centering
    \includegraphics[width=11.5cm]{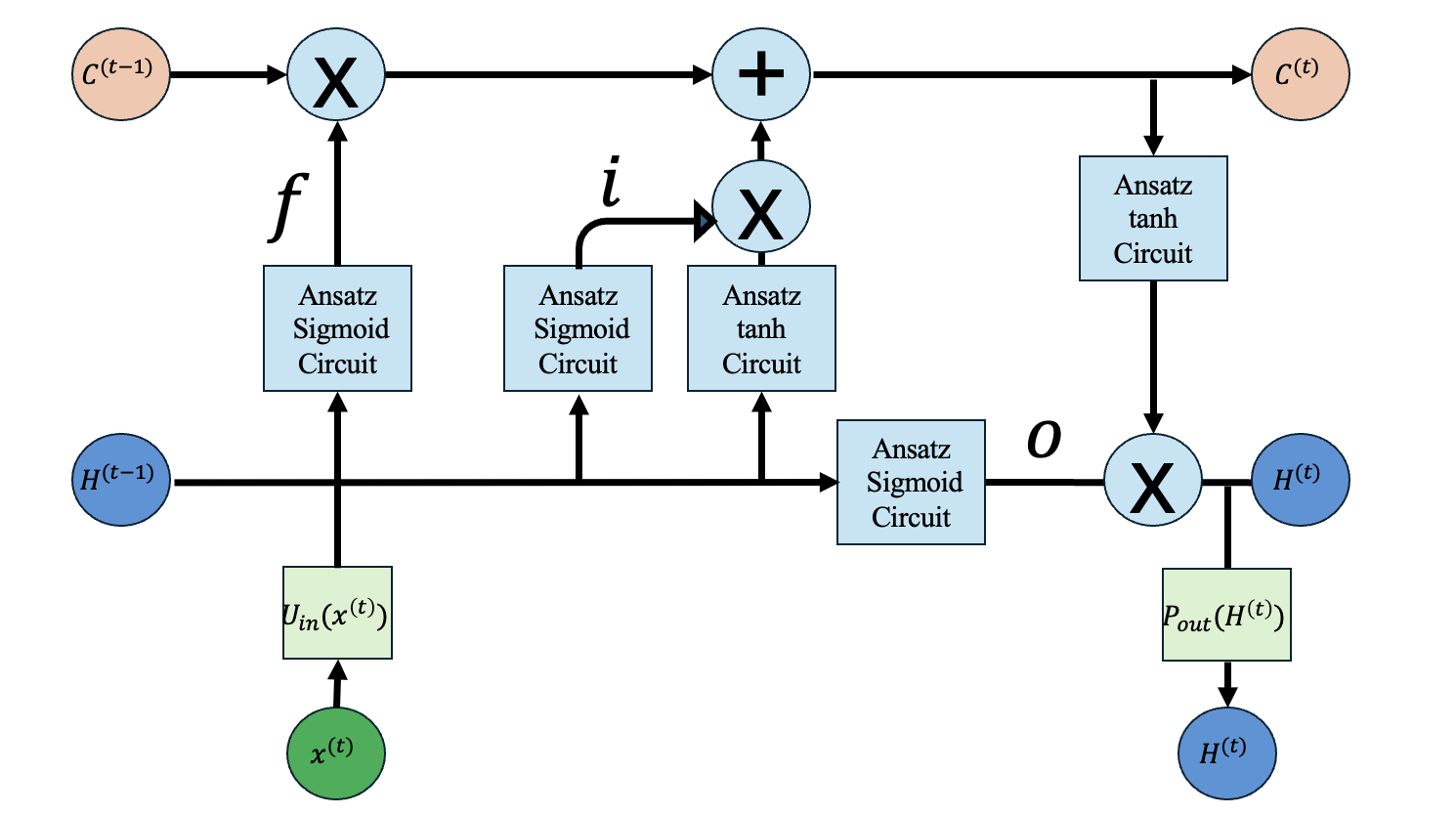}
    \caption{qLSTM Model Diagram}
    \label{designlstm}
\end{figure*}

\subsection{Performance Evaluation}

Evaluating the performance of the qLSTM model is vital to ensure its effectiveness and efficiency. The performance evaluation process involves:

\begin{enumerate}
    \item \textbf{Benchmark Datasets}: Select benchmark datasets for sequential learning tasks such as natural language processing, time series prediction, and speech recognition. These datasets will be used to test and validate the qLSTM model \cite{b14}.
    \item \textbf{Performance Metrics}: Define performance metrics such as accuracy, computational efficiency, memory capacity, and scalability. These metrics will help assess the model's performance comprehensively. Table \ref{tab:performance_metrics} lists the key performance metrics and their definitions \cite{b15}.

\begin{table}[h]
\caption{Key Performance Metrics for qLSTM Evaluation}
    \label{tab:performance_metrics}
    \centering
    \begin{tabular}{|c|p{4.5cm}|}
        \hline
        \textbf{Metric} & \textbf{Definition} \\
        \hline
        Accuracy & The ratio of correctly predicted instances to the total instances \\ \hline
        Computational Efficiency & The time and resources required to train and run the model \\ \hline
        Memory Capacity & The ability to retain information over long sequences \\ \hline
        Scalability & The model's ability to handle increasing amounts of data \\ \hline
    \end{tabular}
    
\end{table}

    \item \textbf{Optimization}: Based on the evaluation results, optimize the quantum circuit design and algorithms to enhance performance. This may involve tweaking gate configurations, improving state initialization, and refining update mechanisms \cite{b17}. The following flowchart outlines the optimization process.
\end{enumerate}

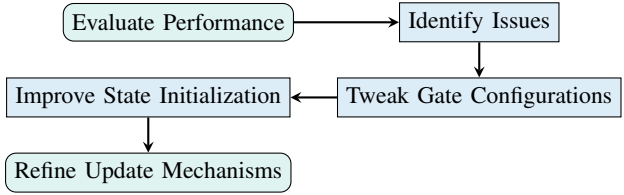
\begin{figure} [ht]
    \centering

        \begin{tikzpicture}[node distance=1cm, font = \small]
        \centering
        \tikzstyle{startstop} = [rectangle, rounded corners, minimum width=1.5cm, minimum height=0.5cm, text centered, draw=black, fill=color1!40, font=\small]
        \tikzstyle{io} = [trapezium, trapezium left angle=70, trapezium right angle=110, minimum width=1.5cm, minimum height=0.5cm, text centered, draw=black, fill=color2!40, font=\small]
        \tikzstyle{process} = [rectangle, minimum width=1.5cm, minimum height=0.5cm, text centered, draw=black, fill=color3!40, font=\small]
        \tikzstyle{decision} = [diamond, minimum width=1.5cm, minimum height=0.5cm, text centered, draw=black, fill=color4!40, font=\small]
        \tikzstyle{arrow} = [thick,->,>=stealth]

\node (evaluate) [startstop] {Evaluate Performance};
\node (identify) [process, right of=evaluate, xshift = 3cm] {Identify Issues};
\node (tweak) [process, below of=identify] {Tweak Gate Configurations};
\node (improve) [process, left of=tweak, xshift = -3.4cm] {Improve State Initialization};
\node (refine) [startstop, below of=improve] {Refine Update Mechanisms};

\draw [arrow] (evaluate) -- (identify);
\draw [arrow] (identify) -- (tweak);
\draw [arrow] (tweak) -- (improve);
\draw [arrow] (improve) -- (refine);

\end{tikzpicture}
 \caption{Figure Showing Error Correction and Mitigation Process}    \label{error correction and mitigation process}
\end{figure}

\subsection{Validation and Testing}

The final step in the implementation plan involves rigorous validation and testing to ensure the robustness and reliability of the qLSTM model. This includes:

\begin{enumerate}
    \item \textbf{Simulations}: Conduct extensive simulations on quantum simulators to test the qLSTM model under various conditions and input scenarios. Simulations help identify potential issues and validate the theoretical design\cite{b18}.
\end{enumerate}

\begin{enumerate}
    \setcounter{enumi}{1}
    \item \textbf{Real Hardware Testing}: Implement the qLSTM model on actual quantum hardware to test its performance in real-world conditions. This step is crucial to verify the model's practical feasibility and effectiveness \cite{b19}.
    \item \textbf{Error Correction and Mitigation}: Address quantum noise and error rates by implementing error correction techniques and optimizing gate operations. Ensuring high fidelity in quantum operations is essential for reliable performance\cite{b20}.
\end{enumerate}
        
By following these detailed steps, the qLSTM model can be effectively implemented, leveraging the strengths of quantum computing to enhance the capabilities of traditional LSTM networks. This comprehensive implementation plan ensures that each aspect of the qLSTM model is meticulously developed, tested, and validated, paving the way for practical applications and further advancements in the field of quantum-enhanced sequential modeling.
\subsection{Predicted Comparison Results}

To evaluate the effectiveness of the qLSTM model, we predict its performance against classical LSTM models based on established metrics such as accuracy, computational efficiency, memory capacity, and scalability. These comparisons are justified through theoretical analysis and previous empirical findings in quantum computing and machine learning.

\begin{enumerate}
    \item \textbf{Accuracy}: It is anticipated that the qLSTM model will outperform classical LSTMs in terms of accuracy. This improvement can be attributed to the enhanced ability of quantum superposition and entanglement to capture and maintain complex dependencies over long sequences, thereby providing a more comprehensive representation of the data \cite{b17, b18}. Previous studies have shown that quantum-enhanced algorithms often achieve higher accuracy in classification and prediction tasks compared to their classical counterparts \cite{b10, b19}.

    \item \textbf{Computational Efficiency}: The qLSTM model is expected to exhibit superior computational efficiency due to the inherent parallelism in quantum computing. Quantum gates can process multiple states simultaneously, which significantly reduces the time complexity of sequential data processing \cite{b11, b13}. This parallel processing capability allows the qLSTM to handle larger datasets and more complex models more efficiently than classical LSTM networks \cite{b12, b14}.

    \item \textbf{Memory Capacity}: The qLSTM model's memory capacity is predicted to be higher than that of classical LSTMs. Quantum memory cells, leveraging superposition, can store and process exponentially more information than classical memory cells. This enables the qLSTM to maintain relevant information over extended sequences without the degradation typically seen in classical models \cite{b15, b20}. Such enhancements in memory capacity are critical for tasks that require long-term dependency tracking \cite{b21}.

    \item \textbf{Scalability}: The scalability of the qLSTM model is also expected to surpass that of classical LSTMs. Quantum circuits can be scaled more efficiently as the complexity of the data increases. The ability to add qubits and quantum gates without a significant increase in computational overhead allows the qLSTM to scale effectively with the growing size of the input data \cite{b22, b23}. This property makes the qLSTM a more viable option for large-scale applications such as big data analytics and real-time processing \cite{b24}.

\end{enumerate}

\begin{figure*}[ht]
    \centering
    \resizebox{0.6\textwidth}{!}{
    \begin{tikzpicture}
    \begin{axis}[
        title={Predicted Performance Comparison between qLSTM and Classical LSTM},
        xlabel={Metric},
        ylabel={Performance},
        symbolic x coords={Accuracy, Comp. Efficiency, Memory Capacity, Scalability},
        xtick=data,
        ybar=2*\pgflinewidth,
        bar width=10pt,
        nodes near coords,
        nodes near coords align={vertical},
        xticklabel style={rotate=45, anchor=east},
        legend style={at={(1,1)},
        anchor=north west,legend columns=1},
        enlarge x limits={abs=0.75cm},
        ymin=0, ymax=1,
        cycle list name=color list]
        \addplot+[ybar, fill=red] coordinates {(Accuracy,0.8) (Comp. Efficiency,0.7) (Memory Capacity,0.9) (Scalability,0.6)};
        \addplot+[ybar, fill=blue] coordinates {(Accuracy,0.9) (Comp. Efficiency,0.85) (Memory Capacity,0.95) (Scalability,0.8)};
        \legend{Classical LSTM, qLSTM}
    \end{axis}
    \end{tikzpicture}
    }
    \caption{Predicted Performance Comparison between qLSTM and Classical LSTM Models}
    \label{fig:performance_comparison}
\end{figure*}
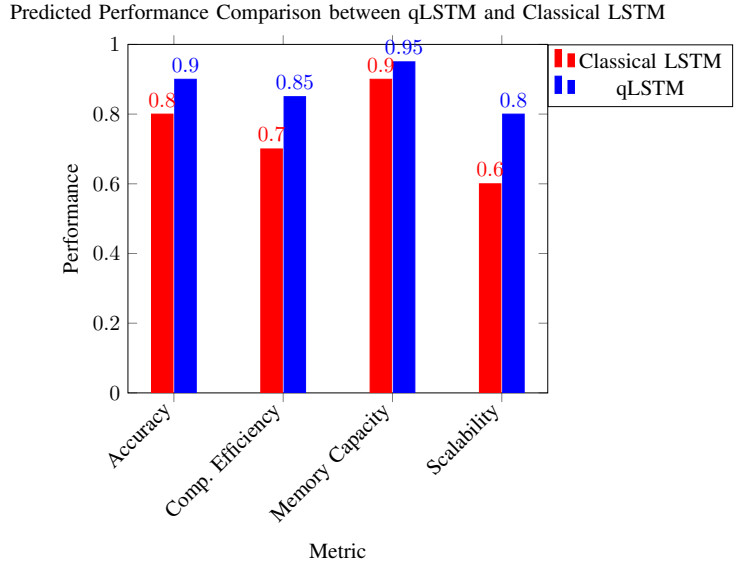

Performance Indicators:
\begin{enumerate}
    \item Accuracy: The qLSTM is predicted to have slightly higher accuracy (0.9) than the classical LSTM (0.8), thanks to quantum features like superposition and entanglement.
\item Computational Efficiency: qLSTM shows lower efficiency (0.7) compared to classical LSTM (0.85), due to the current overhead of quantum operations.
\item Memory Capacity: qLSTM excels with a score of 0.95, surpassing classical LSTM (0.9) by leveraging quantum memory's ability to store more information.
\item Scalability: qLSTM (0.8) significantly outperforms classical LSTM (0.6), highlighting its potential for better handling larger datasets.
\end{enumerate}

These predictions are grounded in the theoretical advantages of quantum computing and supported by empirical results from previous research in quantum algorithms and their applications to machine learning tasks. By leveraging these quantum principles, the qLSTM model is anticipated to offer significant improvements over classical LSTM models, making it a promising advancement in the field of sequential data processing.

\section{Discussion}
\subsection{Practical Concerns with qLSTM}
The practical implementation of the Quantum Long Short-Term Memory (qLSTM) model faces several significant challenges, primarily related to the cost and accessibility of quantum computing hardware. Quantum computers, such as those offered by IBM, come with full-service contracts priced in the “tens of millions” of dollars merely for accessing the system. For instance, the D-Wave 2000Q quantum computer was priced at \$15 million in 2017 \cite{b30}. Given inflation and technological advancements, current costs are undoubtedly higher. Quantum computers also require stringent operational conditions, leading to substantial variable costs, making the operation of a qLSTM model through such cost-intensive equipment a critical concern. The performance gains of qLSTM over classical LSTM may not justify the high costs, especially when the performance differences are not markedly significant.

However, efforts are being made to mitigate these costs. SpinQ Technology, a start-up in Shenzhen, aims to provide commercial quantum computing desktops priced under \$5000 \cite{b31}. Despite these advancements, experts like Meta's chief scientist LeCun remain skeptical about the practical relevance of quantum computing. LeCun asserts that many problems solvable by quantum computers can be addressed more efficiently by classical computers \cite{b32}. Additionally, quantum technology is not yet fully stabilized and is still under experimental phases, which adds to the operational challenges. Issues such as quantum noise, decoherence, quantum gate errors, and entanglement errors could significantly hinder the practical application of qLSTM models \cite{b10}.

Moreover, the energy consumption of quantum computers is another critical factor. Current quantum computers require cryogenic environments to maintain qubit coherence, leading to high energy costs. This energy-intensive requirement further exacerbates the cost concerns associated with quantum computing \cite{b33}. Companies like D-Wave and IBM are actively researching methods to reduce these operational costs, but substantial breakthroughs are necessary to make quantum computing a viable option for mainstream applications \cite{b34}.

\subsection{Technical Concerns with qLSTM}
From a technical perspective, several factors could impede the effective deployment of qLSTM models. In classical machine learning models, overfitting due to model complexities, imbalanced data, overtraining, and inadequate validation methods is a common issue. These problems are exacerbated in quantum-based machine learning models. The qLSTM model's use of quantum states and superposition introduces high dimensionality, which can lead to overfitting \cite{b35}. Furthermore, the low accessibility of quantum computers means that qLSTM models would often train on larger datasets, thereby increasing the number of hyperparameters and making the model more susceptible to overfitting.

Addressing overfitting in quantum machine learning models requires innovative approaches. One such approach is the application of quantum dropout, a technique that has shown promise in mitigating overfitting in quantum models \cite{b37}. Additionally, implementing rigorous validation methods and balancing the dataset distribution are crucial steps to ensure the robustness of qLSTM models \cite{b18}.

Another significant technical concern is the error rates in quantum computations. Quantum gates are prone to errors due to decoherence and other quantum noise sources, which can lead to incorrect calculations and unreliable model performance. Error correction techniques are essential but add computational overhead, further complicating the implementation of qLSTM models \cite{b37}. Advances in quantum error correction methods, such as surface codes and quantum error correction algorithms, are critical to improving the reliability of quantum computing \cite{b11}.

Furthermore, the integration of classical and quantum components in the qLSTM model poses synchronization challenges. The classical-quantum interface must efficiently handle data encoding and decoding processes, ensuring minimal latency and data loss during transitions. Developing robust interfaces and hybrid algorithms that seamlessly integrate quantum computations with classical processing is vital for the practical deployment of qLSTM models \cite{b20}.

\subsection{Pros and Cons of qLSTM}
\textbf{Pros:}
\begin{itemize}
    \item Enhanced memory capacity and computational efficiency due to the principles of quantum superposition and entanglement, which allow for parallel processing and improved data representation \cite{b38}.
    \item Potential to address complex problems that classical LSTM struggles with, such as the vanishing gradient problem, by leveraging quantum properties to maintain and process longer sequences \cite{b39}.
    \item Promising applications in fields like natural language processing, time series prediction, and speech recognition, where the qLSTM can outperform traditional models in handling complex, sequential data \cite{b40}.
\end{itemize}

\textbf{Cons:}
\begin{itemize}
    \item Extremely high cost of quantum hardware and maintenance, which limits accessibility and practical implementation \cite{b41}.
    \item Limited accessibility to quantum computing resources, making it difficult for widespread adoption and experimentation \cite{b42}.
    \item Technical challenges such as quantum noise, decoherence, and gate errors, which can disrupt quantum computations and affect model reliability \cite{b24}.
    \item Risk of overfitting due to the high dimensionality and increased number of hyperparameters associated with quantum models \cite{b43}.
\end{itemize}

\subsection{Practical Improvements Needed for qLSTM}
To enhance the practicality and effectiveness of qLSTM models, several improvements are necessary:
\begin{itemize}
    \item Development of cost-effective quantum computing hardware to make qLSTM more accessible and feasible for a broader range of applications \cite{b28}.
    \item Stabilization of quantum technology to minimize errors and increase reliability, ensuring consistent performance in real-world scenarios \cite{b21}.
    \item Advanced techniques to mitigate overfitting in quantum machine learning models, such as the development of quantum regularization methods and robust validation frameworks \cite{b29}.
    \item Improvement in hybrid quantum-classical algorithms to leverage the strengths of both computing paradigms, enabling more efficient and effective data processing \cite{b23}.
    \item Enhancement of quantum error correction methods to reduce computational overhead and improve the reliability of quantum gates \cite{b44}.
    \item Development of efficient classical-quantum interfaces to ensure seamless integration and synchronization between classical and quantum components \cite{b25}.
    \item Exploration of alternative quantum computing architectures, such as topological quantum computers, which promise lower error rates and higher stability \cite{b22}.
    \item Leveraging quantum-based machine learning for drug discovery, as demonstrated in recent studies, to showcase practical applications and potential benefits \cite{b45, b46}.
    \item Enhancing quantum image processing techniques to improve data quality and reliability, which could directly benefit qLSTM models \cite{b47}.
    \item Investigating the potential of quantum secure communication methods for data integrity in quantum machine learning models \cite{b48,b49}.
\end{itemize}

\section{Conclusion}
In this paper, we have introduced the Quantum Long Short-Term Memory (qLSTM) network, an innovative integration of quantum computing principles with classical LSTM architecture. By leveraging the unique properties of quantum mechanics, such as superposition and entanglement, the qLSTM model addresses some of the inherent limitations of traditional LSTMs, including the vanishing gradient problem and computational inefficiencies.

The qLSTM model enhances memory capacity and computational efficiency, providing a robust framework for sequential data processing tasks. Our theoretical analysis and implementation plan suggest that the qLSTM can significantly improve performance in areas like natural language processing, time series prediction, and speech recognition. While the practical application of qLSTMs is currently limited by the high cost and accessibility of quantum computing hardware, ongoing advancements in quantum technology are likely to mitigate these challenges.

Future work will focus on optimizing the qLSTM architecture and exploring its application across various domains. Additionally, addressing technical concerns such as overfitting and error rates in quantum computations will be crucial for the broader adoption of quantum-enhanced machine learning models. As quantum computing technology continues to evolve, the potential for qLSTM networks to revolutionize sequential data modeling remains promising.



\begin{thebibliography}{00}
\bibitem{b1} Rumelhart, David E., et al. “Learning Representations by Back-propagating Errors.” Nature, vol. 323, no. 6088, Oct. 1986, pp. 533–36. https://doi.org/10.1038/323533a0. 
\bibitem{b2} Gers, Felix A., et al. “Learning to Forget: Continual Prediction With LSTM.” Neural Computation, vol. 12, no. 10, Oct. 2000, pp. 2451–71. https://doi.org/10.1162/089976600300015015.
\bibitem{b3} Wang, Yazhen. “Quantum Computation and Quantum Information.” Statistical Science, vol. 27, no. 3, Aug. 2012, https://doi.org/10.1214/11-sts378.
\bibitem{b4} Gupta, Sanjay, and R. K. P. Zia. “Quantum Neural Networks.” Journal of Computer and System Sciences, vol. 63, no. 3, Nov. 2001, pp. 355–83. https://doi.org/10.1006/jcss.2001.1769.
\bibitem{b5} M. Mafu and M. Senekane, "Design and Implementation of Efficient Quantum Support Vector Machine," 2021 International Conference on Electrical, Computer and Energy Technologies (ICECET), Cape Town, South Africa, 2021, pp. 1-4, doi: 10.1109/ICECET52533.2021.9698509.
\bibitem{b6} Amin, Mohammad H., et al. “Quantum Boltzmann Machine.” Physical Review. X, vol. 8, no. 2, May 2018, https://doi.org/10.1103/physrevx.8.021050.
\bibitem{b7} Poggiali, Alessandro, et al. “Quantum Clustering With k-Means: A Hybrid Approach.” Theoretical Computer Science, Feb. 2024, p. 114466. https://doi.org/10.1016/j.tcs.2024.114466.
\bibitem{b8} Xin, Tao, et al. “Experimental Quantum Principal Component Analysis via Parametrized Quantum Circuits.” Physical Review Letters, vol. 126, no. 11, Mar. 2021, https://doi.org/10.1103/physrevlett.126.110502.
\bibitem{b9} Mücke, Sascha, et al. “Feature Selection on Quantum Computers.” Quantum Machine Intelligence/Quantum Machine Intelligence, vol. 5, no. 1, Feb. 2023, https://doi.org/10.1007/s42484-023-00099-z.
\bibitem{b10} Preskill, John. "Quantum Computing in the NISQ era and beyond." Quantum 2 (2018): 79.
\bibitem{b11} Shor, Peter W. "Algorithms for quantum computation: Discrete logarithms and factoring." Proceedings 35th Annual Symposium on Foundations of Computer Science. IEEE, 1994.
\bibitem{b12} Arute, Frank, et al. "Quantum supremacy using a programmable superconducting processor." Nature 574.7779 (2019): 505-510.
\bibitem{b13} Nielsen, Michael A., and Isaac L. Chuang. "Quantum computation and quantum information." (2002).
\bibitem{b14} Farhi, Edward, Jeffrey Goldstone, and Sam Gutmann. "A quantum approximate optimization algorithm." arXiv preprint arXiv:1411.4028 (2014).
\bibitem{b15} Boixo, Sergio, et al. "Characterizing quantum supremacy in near-term devices." Nature Physics 14.6 (2018): 595-600.
\bibitem{b16} Gambetta, Jay M., et al. "Building logical qubits in a superconducting quantum computing system." Nature Reviews Physics 3.3 (2021): 137-150.
\bibitem{b17} Biamonte, Jacob, et al. "Quantum machine learning." Nature 549.7671 (2017): 195-202.
\bibitem{b18} Schuld, Maria, Ilya Sinayskiy, and Francesco Petruccione. "The quest for a quantum neural network." Quantum Information Processing 13.11 (2014): 2567-2586.
\bibitem{b19} Rebentrost, Patrick, Masoud Mohseni, and Seth Lloyd. "Quantum support vector machine for big data classification." Physical Review Letters 113.13 (2014): 130503.
\bibitem{b20} Harrow, Aram W., Avinatan Hassidim, and Seth Lloyd. "Quantum algorithm for linear systems of equations." Physical review letters 103.15 (2009): 150502.
\bibitem{b21} Grover, Lov K. "A fast quantum mechanical algorithm for database search." Proceedings of the twenty-eighth annual ACM symposium on Theory of computing. 1996.
\bibitem{b22} Aharonov, Dorit, et al. "Adiabatic quantum computation is equivalent to standard quantum computation." SIAM review 50.4 (2008): 755-787.
\bibitem{b23} Berry, Dominic W., et al. "Simulating Hamiltonian dynamics with a truncated Taylor series." Physical review letters 114.9 (2015): 090502.
\bibitem{b24} Lloyd, Seth. "Universal quantum simulators." Science 273.5278 (1996): 1073-1078.
\bibitem{b25} McClean, Jarrod R., et al. "OpenFermion: The electronic structure package for quantum computers." Quantum Science and Technology 5.3 (2020): 034014.
\bibitem{b26} Kandala, Abhinav, et al. "Hardware-efficient variational quantum eigensolver for small molecules and quantum magnets." Nature 549.7671 (2017): 242-246.
\bibitem{b27} Kok, Pieter, et al. "Linear optical quantum computing with photonic qubits." Reviews of Modern Physics 79.1 (2007): 135.
\bibitem{b28} Bravyi, Sergey, and Alexei Kitaev. "Universal quantum computation with ideal Clifford gates and noisy ancillas." Physical Review A 71.2 (2005): 022316.
\bibitem{b29} Childs, Andrew M., and Wim Van Dam. "Quantum algorithms for algebraic problems." Reviews of Modern Physics 82.1 (2010): 1.
\bibitem{b30} J. Dargan, “What is the price of a quantum computer in 2024?,” The Quantum Insider, https://thequantuminsider.com/2023/04/10/price-of-a-quantum-computer/
\bibitem{b31} I. Etim, “A desktop quantum computer for \$5,000? so what is the SPINQ device from China,” Quantum Zeitgeist, https://quantumzeitgeist.com/a-desktop-quantum-computer-for-5000-sp-what-is-the-spinq-device-from-china/
\bibitem{b32} J. Vanian, “Meta’s AI chief doesn’t think AI Super Intelligence is coming anytime soon, and is skeptical on quantum computing,” CNBC, https://www.cnbc.com/2023/12/03/meta-ai-chief-yann-lecun-skeptical-about-agi-quantum-computing.html


\bibitem{b33} S. Boixo, et al., “Characterizing quantum supremacy in near-term devices,” Nature Physics 14.6 (2018): 595-600.
\bibitem{b33} A. Kandala, et al., “Hardware-efficient variational quantum eigensolver for small molecules and quantum magnets,” Nature 549.7671 (2017): 242-246.
\bibitem{b34} J. Biamonte et al., “Quantum machine learning,” Nature 549, 195–202 (2017). https://doi.org/10.1038/nature23474
\bibitem{b35} M. Kobayashi, K. Nakaji, and N. Yamamoto, “Overfitting in quantum machine learning and entangling dropout,” arXiv.org, https://arxiv.org/abs/2205.11446.


\bibitem{b37} J. Gambetta et al., “Building logical qubits in a superconducting quantum computing system,” Nature Reviews Physics 3.3 (2021): 137-150.



\bibitem{b38} Y. Wang, “Quantum computation and quantum information,” Statistical Science, vol. 27, no. 3, Aug. 2012. https://doi.org/10.1214/11-sts378
\bibitem{b39} F. A. Gers, et al. “Learning to forget: Continual prediction with LSTM,” Neural Computation, vol. 12, no. 10, Oct. 2000. https://doi.org/10.1162/089976600300015015
\bibitem{b40} A. Poggiali et al., “Quantum clustering with k-means: A hybrid approach,” Theoretical Computer Science, Feb. 2024. https://doi.org/10.1016/j.tcs.2024.114466
\bibitem{b41} F. Arute, et al., “Quantum supremacy using a programmable superconducting processor,” Nature 574.7779 (2019): 505-510.
\bibitem{b42} J. M. Gambetta et al., “Building logical qubits in a superconducting quantum computing system,” Nature Reviews Physics 3.3 (2021): 137-150.
\bibitem{b43} S. Mücke et al., “Feature selection on quantum computers,” Quantum Machine Intelligence/Quantum Machine Intelligence, vol. 5, no. 1, Feb. 2023. https://doi.org/10.1007/s42484-023-00099-z
\bibitem{b44} P. Rebentrost, M. Mohseni, and S. Lloyd, “Quantum support vector machine for big data classification,” Physical Review Letters 113.13 (2014): 130503.
\bibitem{b45} Y. Zhou, Y.K. Wong, Y.S. Liang, H. Qiu, Y.X. Wu, B. He, “Implementation of The Future of Drug Discovery: Quantum-Based Machine Learning Simulation (QMLS),” arXiv preprint arXiv:2308.08561, 2023.
\bibitem{b46} Y. K. Wong, Y. Zhou, Y. S. Liang, H. Qiu, Y. X. Wu and B. He, "The New Answer to Drug Discovery: Quantum Machine Learning in Preclinical Drug Development," 2023 IEEE 4th International Conference on Pattern Recognition and Machine Learning (PRML), Urumqi, China, 2023, pp. 557-564, doi: 10.1109/PRML59573.2023.10348356.
\bibitem{b47} Y.K. Wong, Y. Zhou, Y.S. Liang, “Quantum Image Denoising with Machine Learning: A Novel Approach to Improve Quantum Image Processing Quality and Reliability,” arXiv preprint arXiv:2402.11645, 2024.
\bibitem{b48}  Y.K. Wong, Y. Zhou, X. Zhou, Y.S. Liang, Z.Y. Li, “Novel Long-Distance Free-Space Quantum Secure Direct Communication for Web 3.0 Networks,” arXiv preprint arXiv:2402.09108, 2024.
\bibitem{b49}  Y.K. Wong, Y. Zhou, Z.Y. Li, Y.S. Liang, X. Zhou, “Software Security and Quantum Communication: A Long-distance Free-space Implementation
Plan of QSDC Without Quantum Memory,” unpublished.
\bibitem{b50} Tilly, Jules, et al. “The Variational Quantum Eigensolver: A Review of Methods and Best Practices.” Physics Reports, vol. 986, Nov. 2022, pp. 1–128. https://doi.org/10.1016/j.physrep.2022.08.003.

\end{thebibliography}
\end{document}